\documentclass[usenatbib,hyperref]{mn2e}
\usepackage{times}
\usepackage{epsfig}
\usepackage{amssymb,amsmath}
\usepackage{aas_macros}
\title[Disentangling satellite galaxy populations]{Disentangling satellite galaxy populations using orbit tracking in simulations}
\author[K. A. Oman, M. J. Hudson \& P. S. Behroozi]{\newauthor Kyle A. Oman$^{1,}$\thanks{kaoman@uwaterloo.ca}, Michael J. Hudson$^{1,2}$, Peter S. Behroozi$^{3}$
\\
$^{1}$ Department of Physics and Astronomy, University of Waterloo, Waterloo, Ontario, N2L 3G1, Canada\\
$^{2}$ Perimeter Institute for Theoretical Physics, Waterloo, Ontario, N2L 2Y5, Canada\\
$^{3}$ Kavli Institute for Particle Astrophysics and Cosmology; Physics Department, Stanford University; Department of Particle Physics and\\~~Astrophysics, SLAC National Accelerator Laboratory, Stanford, CA 94305\\
}

\date{\today}

\def\Msun{\hbox{$\rm\thinspace M_{\odot}$}}

\begin{document}
\label{firstpage}
\maketitle

\begin{abstract}
Physical processes regulating star formation in satellite galaxies represent an area of ongoing research, but the projected nature of observed coordinates makes separating different populations of satellites (with different processes at work) difficult. The orbital history of a satellite galaxy leads to its present-day phase space coordinates; we can also work backwards and use these coordinates to statistically infer information about the orbital history. We use merger trees from the MultiDark Run 1 N-body simulation to compile a catalogue of the orbits of satellite haloes in cluster environments. We parametrize the orbital history by the time since crossing within 2.5 $r_{\textrm{vir}}$ of the cluster centre and use our catalogue to estimate the probability density over a range of this parameter given a set of present-day projected (i.e. observable) phase space coordinates. We show that different populations of satellite haloes, e.g.\ infalling, backsplash and virialized, occupy distinct regions of phase space, and semi-distinct regions of projected phase space. This will allow us to probabilistically determine the time since infall of a large sample of observed satellite galaxies, and ultimately to study the effect of orbital history on star formation history (the topic of a future paper). We test the accuracy of our method and find that we can reliably recover this time within $\pm 2.58$~Gyr in 68 per cent of cases by using all available phase space coordinate information, compared to $\pm 2.64$~Gyr using only position coordinates and $\pm 3.10$~Gyr guessing `blindly', i.e. using no coordinate information, but with knowledge of the overall distribution of infall times. In some regions of phase space, the accuracy of the infall time estimate improves to $\pm 1.85$~Gyr. Although we focus on time since infall, our method is easily generalizable to other orbital parameters (e.g.\ pericentric distance and time).
\end{abstract}

\section{Introduction}\label{sec-introduction}

We know that galaxies in clusters are typically more `red and dead' than their counterparts in the field \citep{2004ApJ...601L..29H,2004ApJ...615L.101B}, as well as dominantly elliptical \citep[rather than spiral, see][]{1980ApJ...236..351D}. This is thought to be due to a mechanism or combination of mechanisms that halt the collapse of cold gas into stars in a satellite galaxy as it orbits within a deep potential well -- either by heating or removing the gas or by preventing the cooling of additional gas and consuming the existing supply. Some renowned mechanisms include ram pressure stripping \citep[e.g.\ ][]{1972ApJ...176....1G,1999MNRAS.308..947A,2007A&A...472....5J,2010MNRAS.408.1417S}, tidal stripping \citep{2006MNRAS.369.1021M}, harassment \citep{1996Natur.379..613M,2010MNRAS.405.1723S}, strangulation \citep{1980ApJ...237..692L,2000ApJ...540..113B} and mergers \citep{1972ApJ...178..623T,2006MNRAS.373.1013C}. Internal quenching mechanisms have also been proposed -- e.g.\ shock heating \citep{2003MNRAS.345..349B,2006MNRAS.368....2D}, AGN heating \citep{2006MNRAS.365...11C,2006ApJ...648..164M}, see also \citet{2010MNRAS.407..749G} -- which, while unable to account for the difference between field and cluster galaxies, may each contribute to increasing the red fraction in both environments.

It has so far been found difficult to produce a semi-analytic model for quenching that both reproduces the observed star formation rate (SFR) distribution in clusters (\citet{2012arXiv1206.3571W}, but see \citet{2010MNRAS.406.2249W} for some recent improvements). However, all the environmental quenching mechanisms listed above have at least one common characteristic : each is sensitive to the path taken through the cluster. Tidal stripping is strongest near the cluster centre. Ram pressure stripping varies with the local density of hot cluster gas and the relative velocity of the satellite, both of which vary radially in the cluster. Mergers are more probably in the outskirts of clusters. Harassment is most effective during high speed encounters, which occur near the cluster core. Strangulation is triggered by the removal of the halo gas of the satellite, obsensibly via one of the aforementioned mechanisms. An environmental quenching model can therefore reasonably be expected to depend on the previous positions and velocities, or more concisely the orbital history, of satellite galaxies.

The orbital history of a galaxy is not directly observable; if we hope to compare environmental quenching models to observational data, we need some way to characterize the orbit of an observed galaxy. We might parametrize the orbital history of a galaxy in any of a number of ways -- for instance by the time since infall into the cluster potential, the distance of closest approach to the cluster centre or the number of orbits completed since infall -- then attempt to find correlations between these parameters and observables using N-body simulation data. The observables we consider available are the distance between the satellite and the cluster centre projected on to the plane of the sky and the component of the velocity of the satellite relative to the cluster centre along the line of sight (LoS). In some nearby clusters it may also be possible to obtain distances to the cluster centre and the satellite via direct distance measurements (e.g.\ Tully-Fisher relation, surface brightness fluctuations, SNIa luminosities, etc.), but in general these will be much less precise than the other two coordinates -- we will consider this information inaccessible. 

Several authors \citep{2004MNRAS.355..819G,2011MNRAS.413.1973W,2012MNRAS.423.1277D,2012arXiv1211.3411T} have shown that the present radial distance from the cluster centre is negatively correlated to the time since infall into the cluster and other orbital parameters. \citet{2005MNRAS.356.1327G} were amongst the first to compare the velocity distributions of different satellite populations \citep[see also][]{2005MNRAS.364..424W}, and more recently \citet{2011MNRAS.416.2882M} presented a method for separating satellite populations based on, amongst other parameters, their LoS velocities.

This paper presents our method for reconstructing the parametrized orbital history of a satellite galaxy based on its present-day observable phase space coordinates; we defer its applications to future papers. Here, we focus on one parameter -- the time since infall on to a cluster-sized host -- but stress that our method is easily adapted to other parameter choices. The time since the satellite first passed pericentre on its orbit in the cluster and the distance of closest approach to the cluster centre are two examples of other parameter choices that will be of interest in future applications of our method.

In section~\ref{sec-data} we describe the N-body simulation data and its conversion into halo catalogues and merger trees. In section~\ref{sec-method} we present the method used convert the merger trees into our catalogues of satellite orbits, and to a probability distribution function of times since infall given a pair of coordinates in phase space. In section~\ref{sec-discussion} we discuss the importance of LoS velocity data in discriminating between different populations of satellites. We summarize in section~\ref{sec-conclusions}.

We assume the same cosmology used in the Bolshoi and Multidark Run 1 simulations with $\Omega_m=0.27$, $\Omega_\Lambda=0.73$, $\Omega_b=0.0469$, $n_s=0.95$, $h_0=0.70$, $\sigma_8=0.82$ \citep{2012MNRAS.423.3018P}.

\section{Simulations}\label{sec-data}

To obtain a large sample of satellite orbits, we use the output of the MultiDark Run 1 (MDR1) dark matter only simulation with 1 Gpch$^{-1}$ box side length, $2048^3$ particles, current cosmology (WMAP5/WMAP7), $8.7 \times 10^9 h^{-1}$~\Msun~mass resolution and 7 kpc force resolution. The simulation was run from $z=65$ to the present; the majority of snapshots are output at even intervals in scale factor $a$ with some irregular intervals at small $a$. The resolution in scale factor is of 0.0304 before $a\sim0.7$ ($z\sim0.43$) and doubles to 0.0152 afterwards (corresponding to a time resolution of about 0.210~Gyr at $z=0$). This is taken into account in our discussion below. Full details of the simulation parameters are described in \citet{2012MNRAS.423.3018P}. The simulation snapshots were processed by the \textsc{rockstar} halo finder \citep{2011arXiv1110.4372B} and the merger tree code presented in \citet{2011arXiv1110.4370B}. To produce the data used in this paper, the merger tree code was slightly modified so that a halo may contain satellite haloes at distances of up to 2.5 times its virial radius ($r_{\textrm{vir}} = r_{360\textrm{b}}$) from its centre; the default code finds satellites within only 1.0 virial radii. This allows us to track satellites out to the largest apocentric distances which \citet{2004A&A...414..445M} \citep[see also][]{2005MNRAS.356.1327G,2000ApJ...540..113B,2009ApJ...692..931L} show to be about $\sim 2.5 r_{200\textrm{c}} (\sim 2.2 r_{360\textrm{b}})$ by using host-satellite linking as a proxy for cluster membership.

In the following discussion, we will denote full 6D cluster-centric coordinates $(r,v)$. The position and velocity centre of a halo is determined by averaging the positions and velocities of the subset of halo particles which minimizes the expected Poisson error in the coordinates, i.e. the particles occupying the region of highest local density \citep[for more detail on how the coordinates are determined by the halo finder, we refer the interested reader to][\S 3.5.1]{2011arXiv1110.4372B}. 

Projected coordinates will be denoted $(R,V)$. Projection is done along the arbitrarily chosen third ($z$-)axis of the simulation box and includes a correction to the velocities to account for the Hubble flow. With this correction, the simulated velocity differences can be directly compared to observation data, where velocity differences would presumably be measured using a redshift difference. The projected distance between two points is $R_{12}=\sqrt{(r_{2,x}-r_{1,x})^2+(r_{2,y}-r_{1,y})^2}$ and the relative velocity of point 2 with respect to point 1 is $V_{12}=|(v_{2,z}-v_{1,z}) + H(r_{2,z}-r_{1,z})|$. Note that $V\geq0$; this encodes our assumption that the distances to the two points are not known precisely, so only the magnitude of their relative velocity can be determined.

For ease of comparison between satellites of different hosts, all spatial coordinates are normalized to the virial radius $r_{\textrm{vir}}$ of the host halo, which is defined using the formula of \citet{1998ApJ...495...80B}: the radius enclosing a region with an overdensity of 360 times the background density at $z=0$. For readers more accustomed to normalization by $r_{200\textrm{c}}$, an approximate conversion at $z=0$ is $\frac{r_{200\textrm{c}}}{r_{\textrm{vir}}}\sim1.13$. All velocity coordinates are normalized to the rms velocity dispersion $\sigma$ of the host halo, measured in 3D.

\section{Method}\label{sec-method}

\begin{figure*}
\leavevmode \epsfxsize=\columnwidth \epsfbox{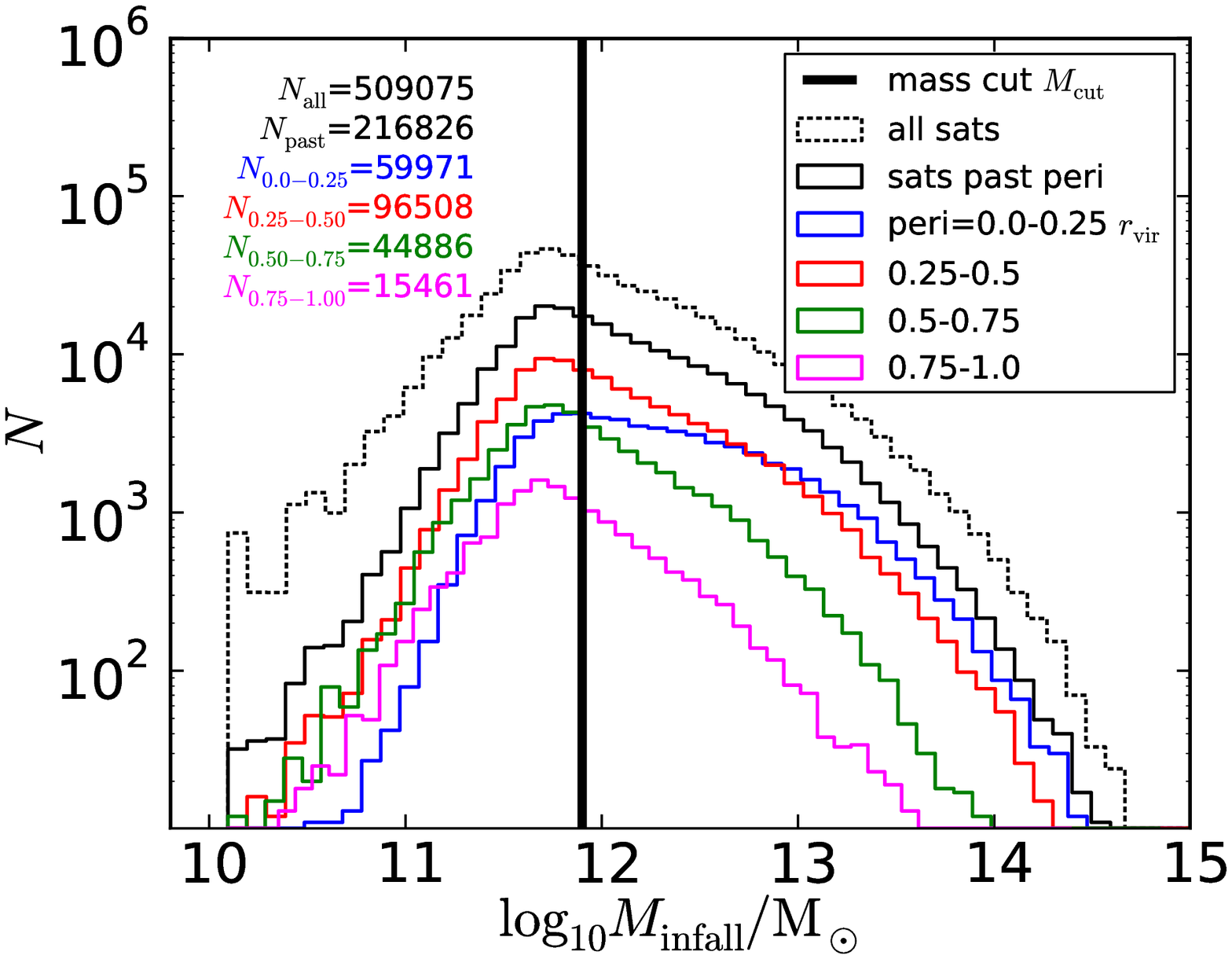}
\hfill
\leavevmode \epsfxsize=\columnwidth \epsfbox{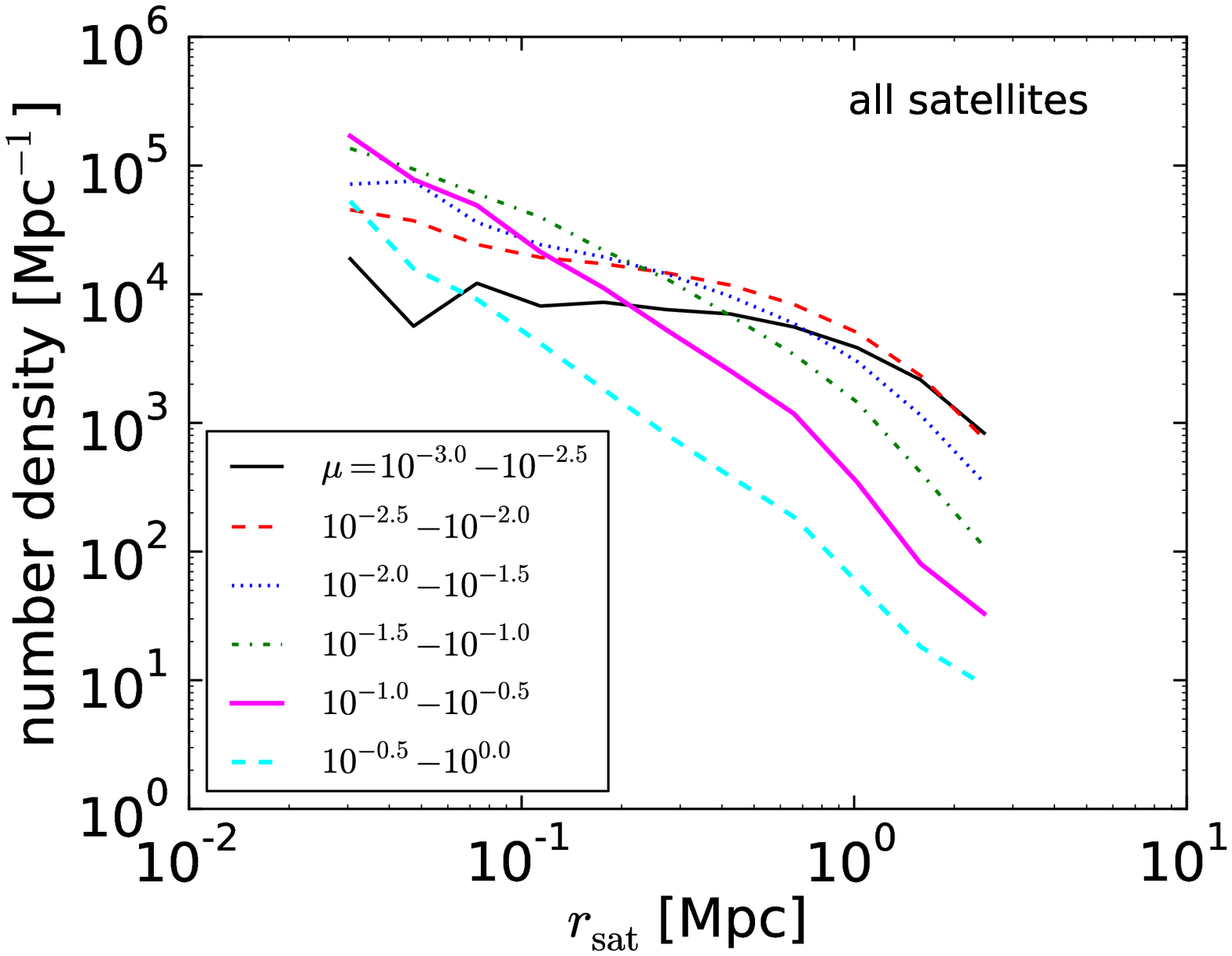}
\caption{These two figures provide a measure of the importance of resolution effects in our analysis; see section \ref{sec-method} and \S\ref{subsec-resolution} for discussion. \emph{Left panel:} The mass function of all satellite haloes in our initial sample (dotted black line) and the mass function of satellite haloes that have experienced at least one pericentric passage (solid black line), where in both cases the masses are measured at the time of infall on to their host. The coloured lines separate the satellites that have experienced a pericentric passage into bins of pericentric distance, as labelled. The total number of satellites contributing to each mass function are labelled $N$. We impose a mass cut as indicated by the thick vertical line ($M_{\textrm{cut}}=10^{11.9}$~\Msun), yielding a sample which is as complete as reasonably possible above $M_{\textrm{cut}}$. \emph{Right panel:} Number density as a function of satellite radial position for various mass ratios $\mu$, where the mass of the satellite is measured at the time of infall. The host halos have masses $\geq 10^{14}$~\Msun. Satellites with low mass ratios are underabundant toward the centre of the cluster environment.\label{mass_function}}
\end{figure*}

First, the orbital history of a satellite and its host at $z=0$ are determined. Then the infall time of the satellite into that host halo is defined as the earliest time at which the satellite's progenitor identifies the $z=0$ host's progenitor as its host. A host-satellite pair may be identified as soon as the satellite is within 2.5 times the virial radius of the host; though other criteria must still be met, as outlined in \citet{2011arXiv1110.4370B}, in practice these are usually met immediately when the satellite crosses within $2.5 r_{\textrm{vir}}$. Only the orbital history with respect to the final ($z=0$) host is considered, and in cases where a satellite has a hierarchy of hosts, intermediate hosts are ignored and the satellite identifies the largest as its host; in other words, the present work ignores ``group pre-processing''. Hosts are selected to be ``cluster-sized'', which we define as a halo mass $>10^{14}$~\Msun. With the MDR1 dataset, this yields a catalogue of $\sim570,000$ satellite orbits belonging to $\sim24,500$ different hosts. 

To ensure a sample complete in satellite mass, and that we are minimally sensitive to artificial disruption of satellite haloes, we impose a mass cut at $M_{\textrm{cut}} = 10^{11.9}$~\Msun, where the mass is measured \emph{at the time of infall}. Using the stellar-to-halo mass ratios outlined in \citet{2012arXiv1207.6105B}, this corresponds to a cut in stellar mass at $\sim10^{10.3}$~\Msun. The left panel of Fig.~\ref{mass_function} shows the mass function for all satellites, and for satellites that have experienced at least one pericentric passage in bins of proximity of pericentric passage. Our mass cut is safely above the mass resolution limit of the simulation. As in essentially all cosmological simulations, MDR1 halos experience artifical disruption in high density environments \citep{2008MNRAS.391.1489K,1999ApJ...522...82K}. This is the reason for the underabundance of lower mass satellites that have experienced a close approach to a cluster centre (blue curve in Fig.~\ref{mass_function}, left panel). From these mass functions, we estimate that less than 20\% of halos with masses above our mass cut and pericentric distances in the range $0.0-0.25 r_{\textrm{vir}}$ have been artificially disrupted. Most of the artificially disrupted satellites are in the mass range $10^{11.9}-10^{13.1}$~\Msun. We estimate that these missing satellites account for less than 4\% of the total halo population above our mass cut. The right panel of Fig.~\ref{mass_function} shows the number density of satellite halos as a function of radial position for bins of satellite mass relative to the host halo. Observations constrain the slope of this power law relation to be between $-1.7$ and $-1.5$, regardless of mass ratio \citep{2012ApJ...745...16T}. The slopes shown in Fig~\ref{mass_function} are somewhat steeper at about $-1.9$ for mass ratio $\mu=10^{-0.5}-10^{0}$. The reason for this steeper slope is not precisely known, but the higher resolution Bolshoi simulation exhibits the same slope of $-1.9$ so we surmise that it is not due to a resolution effect. The key feature that we wish to highlight is that the slope (and shape) of this relationship is mass ratio \emph{dependent} in our dataset, with satellites that are smaller relative to their host being less abundant closer to the centre of the host. The underabundances of satellites highlighted by each panel of Fig. \ref{mass_function} are due to the same population of satellites; those which have orbited to within $\lesssim0.25r _{\textrm{vir}}$ of their host and have a mass $\lesssim 10^{13.1}$~\Msun. See \S\ref{subsec-resolution} for a discussion of the impact of these missing satellites on our results.

We impose one final cut, removing satellites that existed for less than 3 simulation snapshots before falling into a cluster. This prevents haloes near the mass resolution limit of the simulation from appearing suddenly inside a cluster and being assigned meaningless infall times. The remaining 242,790 satellites were binned in 100 projected position bins in $0.0 \leq R/r_{\textrm{vir}} \leq 2.5$ and 100 projected velocity bins in $0.0 \leq V/\sigma \leq 2.0$. The set of infall times in each bin was used to create a probability distribution function (PDF) of infall times for each bin.

\section{Results}\label{sec-discussion}

Before applying our PDFs to modelling environmental quenching (which will be the focus of a future paper), we should have an understanding of a few systematic effects inherent in the method presented in section~\ref{sec-method}. We will discuss the impact of projecting the data in both the radial and velocity coordinates in~\S\ref{subsec-projection}. In~\S\ref{subsec-pdfs} we will present the PDFs and discuss some of their features. In~\S\ref{subsec-mass} we discuss the effects of both host and satellite mass on the distribution of satellites in phase space. Finally, in~\S\ref{subsec-resolution} we discuss the impact of resolution effects on our results.

\subsection{Projection effects}\label{subsec-projection}

While a simulation provides accurate values for all six phase space coordinates of an object, a typical astronomical observation can only measure three. The right ascension and declination give two spatial coordinates (the distance is unknown), while comparing the spectra of two objects can give the difference in velocity between them, but only for motion in a direction along the LoS. Since we ultimately wish to infer properties of observed objects from their coordinates, we must restrict our knowledge of simulated objects to these same coordinates. This can be achieved by ignoring one of the spatial coordinates of the simulation box -- in our case, the third -- and considering only the velocity coordinate corresponding to the ignored spatial coordinate. Additionally, we include a correction to the projected velocity to account for the Hubble flow. This transformation applied to the spatial coordinates of a point $\mathbf{x}$ and its velocity $\mathbf{u}$ relative to a reference point $\mathbf{y}$ and its velocity $\mathbf{w}$ can be expressed:
\begin{align*}
\mathbf{r} = (\mathbf{x}-\mathbf{y}) &= (x_1-y_1,x_2-y_2,x_3-y_3) \\
\Rightarrow \mathbf{R} &= (x_1-y_1,x_2-y_2)\\
\mathbf{v} = (\mathbf{u}-\mathbf{w}) &= (u_1-w_1, u_2-w_2, u_3-w_3) \\
\Rightarrow \mathbf{V} &= (\left|(u_3-w_3) + H(x_3-y_3)\right|)
\end{align*}
We predict the effect of projection on the radial coordinate by considering a random uniform distribution of points on a spherical shell. For this distribution the relationship between the actual ($r$) and projected ($R$) radial coordinates is characterized by $\langle\frac{R}{r}\rangle = \frac{\pi}{4}\pm\sqrt{\frac{2}{3}-\frac{\pi^2}{16}}\sim0.79\pm0.22$ (1$\sigma$ scatter). In our sample of satellite galaxies, we find that $\langle\frac{R}{r}\rangle$ agrees with this prediction to within 1\%, both in the mean and in the scatter. The projected radial coordinate tracks the 3D radial coordinate more closely and more consistently at larger projected radii; at $R\sim r_{\textrm{vir}}$, $\langle\frac{R}{r}\rangle \sim 0.83\pm0.17$ while at $R\sim 0.1 r_{\textrm{vir}}$, $\langle\frac{R}{r}\rangle \sim 0.51\pm0.31$. The observed velocity coordinate does not have as straightforward a relationship with the actual quantity of interest -- the radial component of the velocity difference between two points -- but the information about two components is lost instead of one, so we expect a much larger typical difference between projected velocity and true velocity. We also lose the sign of the one remaining component of velocity since, without knowledge of the distances to the two points, we cannot know whether the distance between them is increasing or decreasing.

\begin{figure*}
\leavevmode \epsfxsize=\columnwidth \epsfbox{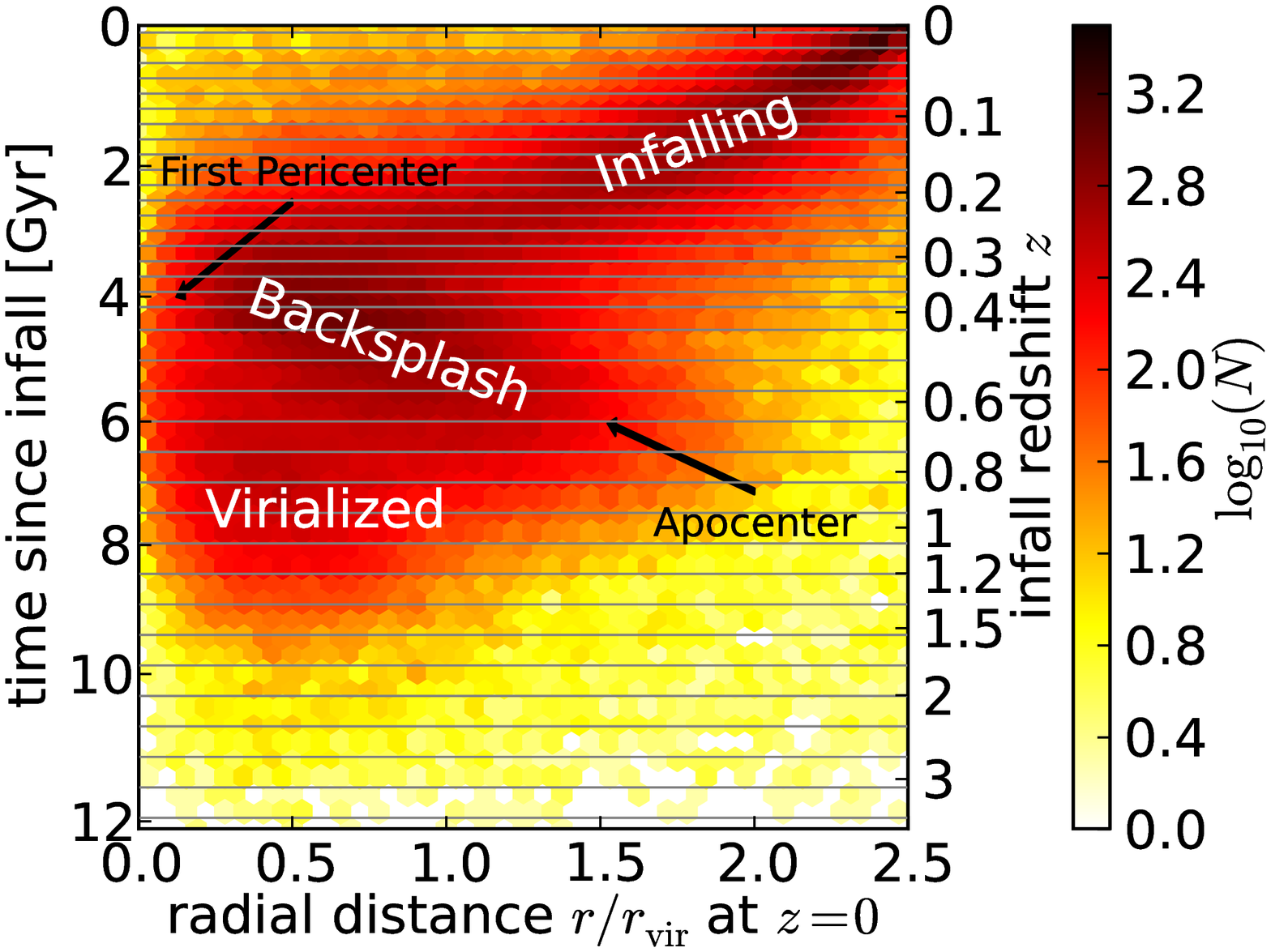}
\hfill
\leavevmode \epsfxsize=\columnwidth \epsfbox{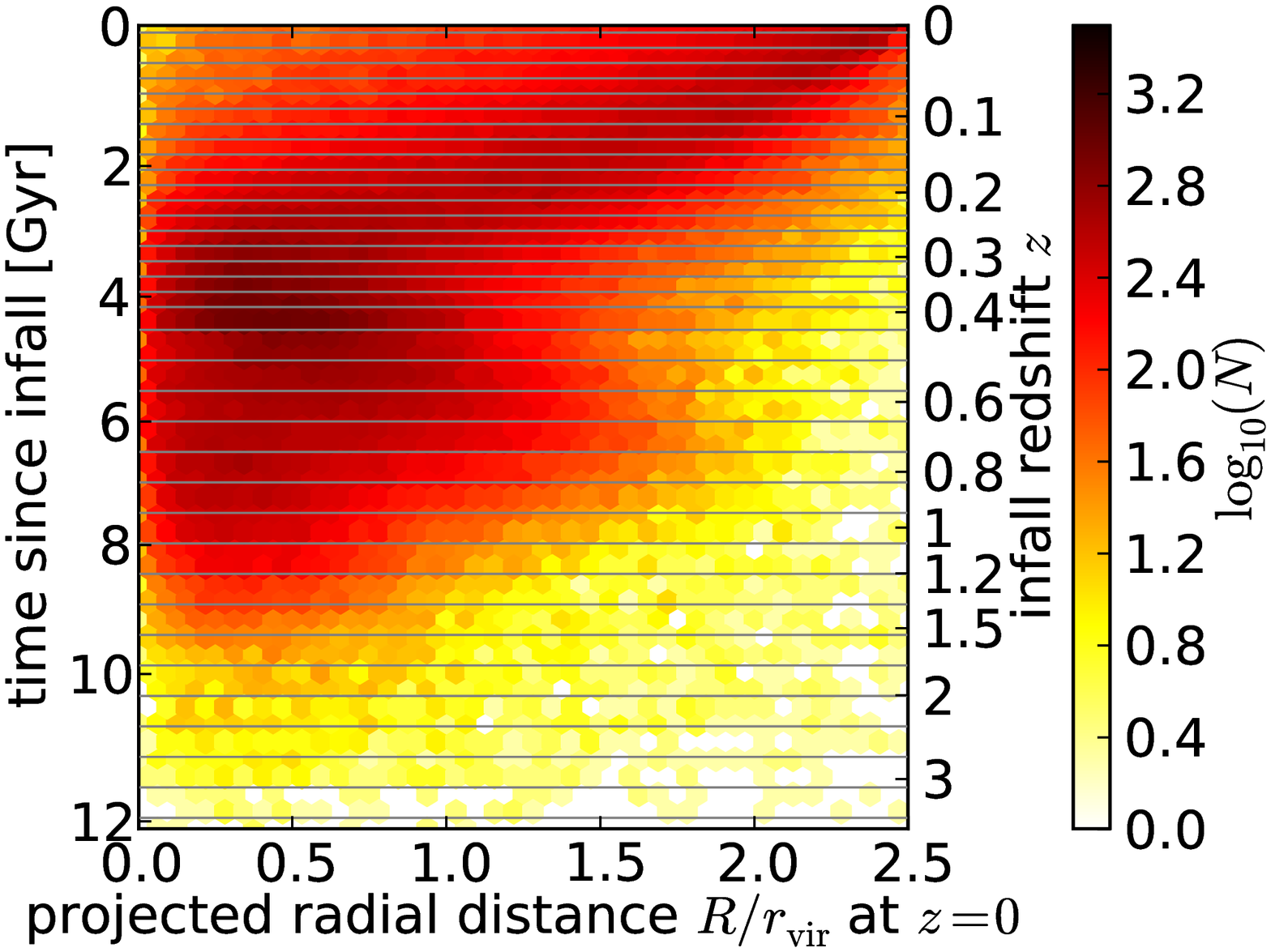}
\caption{\emph{Left panel:} Abundance of satellites in bins of cluster infall time, defined as first inward crossing of 2.5~$r_{\textrm{vir}}$, and radial distance from cluster centre at $t=0$. In the interval $t=0$ to $t\sim3.5$ the `infalling' population is visible, composed of satellites that have not yet experienced a pericentre passage. From $t\sim3.5$ to $t\sim6$ the `backsplash' population is visible, composed of satellites that have passed pericentre once and are approaching apocentre ($t\sim6$). The `virialized' population ($t>6$) are on a second or subsequent orbit. Gray horizontal lines illustrate the times of simulation snapshots. \emph{Right panel:} same as left panel, but using radial distance from cluster centre projected along one axis of the simulation box (simulating coordinates accessible in observations). This widens the distribution of radii of satellites at a given infall time, but the same populations as in the left panel are visible.\label{infall}}
\end{figure*}

The left panel of Fig.~\ref{infall} shows the distribution of satellite infall times\footnote{The infall times produced by the method of section~\ref{sec-method} occur at discrete times -- those of the simulation snapshots. For the purposes of visualization only, some scatter was added to the times.} as a function of radial distance from the host centre \emph{measured at $z=0$} before projection. Satellites that have recent infall times (near the top of the diagram) are necessarily concentrated near the edge of our definition of the cluster at 2.5~$r_{\textrm{vir}}$; they have not had enough time to move anywhere else. A typical crossing time for our sample of clusters is about 6-8~Gyr, so most satellites that fell in 3-4~Gyr ago are near the centre of the host. Note that this seemingly long crossing time is due to our definition of the edge of the cluster at 2.5~$r_{\textrm{vir}}$; the typical time to cross from $r_{\textrm{vir}}$ to pericentre and back to $r_{\textrm{vir}}$ is about 2~Gyr. The satellite also spends a significant amount of time outside the virial radius after this initial crossing; it takes a further $\sim$1~Gyr for a typical satellite to reach apocentre after making its first outbound crossing of $r_{\textrm{vir}}$.

There is a spread in the time taken to reach pericentre and the radii of the pericentres caused by the variety of possible orbits and details of the host potentials. The first apocentre after infall typically occurs after about 6~Gyr; the population of objects between first pericentre and apocentre is termed `backsplash'. Satellites with infall times earlier than $\sim 7$~Gyr have less distinct features in their distribution due to the increasing impact of variations in orbital history, but the majority are confined within $\sim 1$~$r_{\textrm{vir}}$; we call this population `virialized'. We note an overall decreasing number of satellites with increasing time since infall. Some satellites with early infall times are disrupted by tidal interactions and do not appear in our merger trees. In other cases two infalling satellites may merge and appear in the trees as a single satellite. Finally, some halos may not host a galaxy. These three effects mean that observations of satellite galaxies around a cluster do not correlate perfectly with the distribution of dark matter halos around that cluster; this needs to be taken into account when applying our method to a sample of observed galaxies, but does not impact the method itself. There is one other contribution to the decrease in number of satellites with increasing time since infall. Because our mass limit of $10^{11.9}$~\Msun is relatively large, at early times haloes above this limit were somewhat rarer.

The right panel of Fig.~\ref{infall} shows the effect of projection in the radial coordinates on the same distribution as in the left panel. Features are shifted somewhat to lower radii (consistently with our expectation of $\langle\frac{R}{r}\rangle = \frac{\pi}{4}$) and broadened by the scatter about this mean deformation. All the populations and features discussed above are still identifiable. One notable omission from this diagram is any foreground/background objects, which are common in observational samples, that could be confused with the satellite populations.

There is one other important effect to consider when interpreting Fig.~\ref{infall} (and others involving infall times). As a host halo accretes mass, its virial radius grows (slowly, except in the case of major mergers). Because of this, an orbiting satellite may appear to move further in coordinates scaled to $r_{\textrm{vir}}$ than it otherwise would as the coordinates grow around it. This does not have a large impact on the positions, but contributes some of the scatter in the radial coordinate. One might consider choosing some radius that is constant with time to scale each halo, but other choices, such as the virial radius at the last snapshot, also introduce similar effects.

\begin{figure*}
\leavevmode \epsfxsize=2\columnwidth \epsfbox{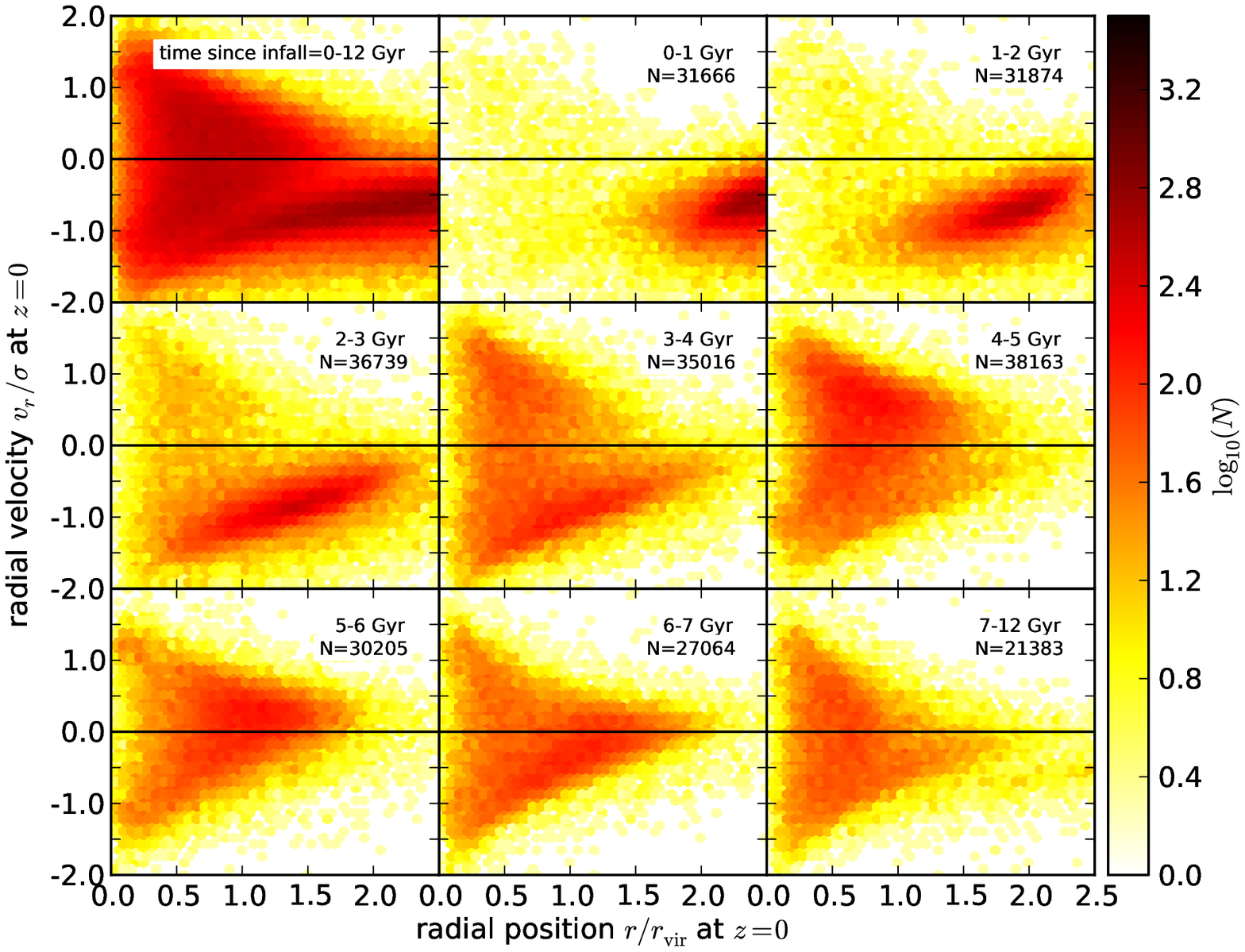}
\caption{The upper left panel shows the phase space distribution of satellite haloes (no projection). The infalling population of haloes is particularly distinct, forming the long dark bar with $v<0$. The backsplash population forms the upper and lower edges of the rest of the main distribution, while the virialized population fills in the centre. The other panels correspond to bins by satellite infall time, as labelled (most bins are 1~Gyr). Each panel shows the $z=0$ distribution of satellites in phase space. Different satellite populations occupy distinct regions of phase space; compare for instance the upper right panel showing mostly infalling satellites, the centre right panel showing mostly backsplash satellites and the lower right panel showing mostly virialized satellites. Each bin is also labelled with the number of satellites $N$ contained in the bin.\label{infall_map_binned}}
\end{figure*}

Next, we consider the effect of projection on the velocity coordinates. The upper left panel of Fig.~\ref{infall_map_binned} shows the distribution of satellite haloes in phase space at $z=0$. A typical halo would, given enough time, progress from large radii and low velocities down to low radii and high negative velocities, then switch to high positive velocity as it passes pericentre. From there it follows a series of progressively shrinking concentric semicircles or chevrons, jumping from negative to positive velocity at each pericentric passage \citep[for a more in depth theoretical background, see][especially fig. 6 therein]{1985ApJS...58...39B}. This normal progression is well represented by the distribution of our halo sample, however the individual orbital `shells' are not visible since they overlap, and we only expect 1-2 shells given the orbital timescales and ages of these systems. Also, close encounters redistribute some haloes off this idealized track.

\begin{figure*}
\leavevmode \epsfxsize=2\columnwidth \epsfbox{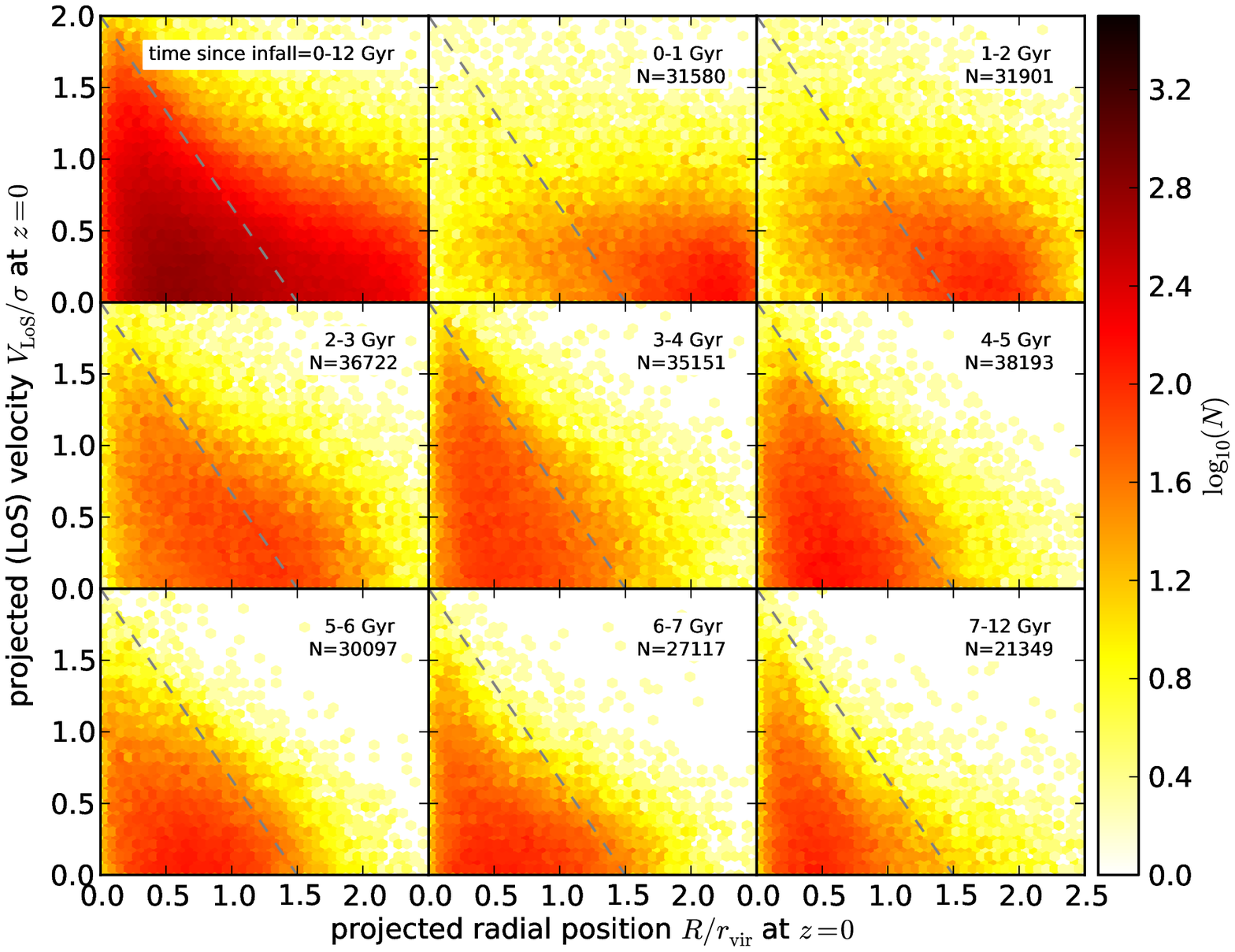}
\caption{As in Fig.~\ref{infall_map_binned}, but using radial distance from cluster centre projected along one axis of the simulation box and velocity projected along the same axis (again simulating coordinates of an observed system). Much of the structure visible in Fig.~\ref{infall_map_binned} is lost in projection, but the use of LoS velocity data allows better spearation of different satellite populations than using the radial coordinate alone. The dashed gray line marks $\frac{V}{\sigma}=-\frac{4}{3}\frac{R}{r_{\textrm{vir}}}+2$; see \S 4.2. \label{infall_map_proj_binned}}
\end{figure*}

Based on this expected movement through phase space in time, satellite haloes with different infall times should occupy different regions of phase space. This is shown in Fig.~\ref{infall_map_binned}, where the phase space distribution of haloes is plotted for a variety of bins in infall time. Fig.~\ref{infall_map_proj_binned} shows the same binned distributions, but in projected coordinates, simultaneously showing the effects of both radial and velocity projections, but emphasizing the latter. Much of the structure visible in Fig.~\ref{infall_map_binned} is lost in projection, but haloes with different infall times still occupy different regions of phase space.

\subsection{Infall time PDFs}\label{subsec-pdfs}

\begin{figure*}
\leavevmode \epsfxsize=2\columnwidth \epsfbox{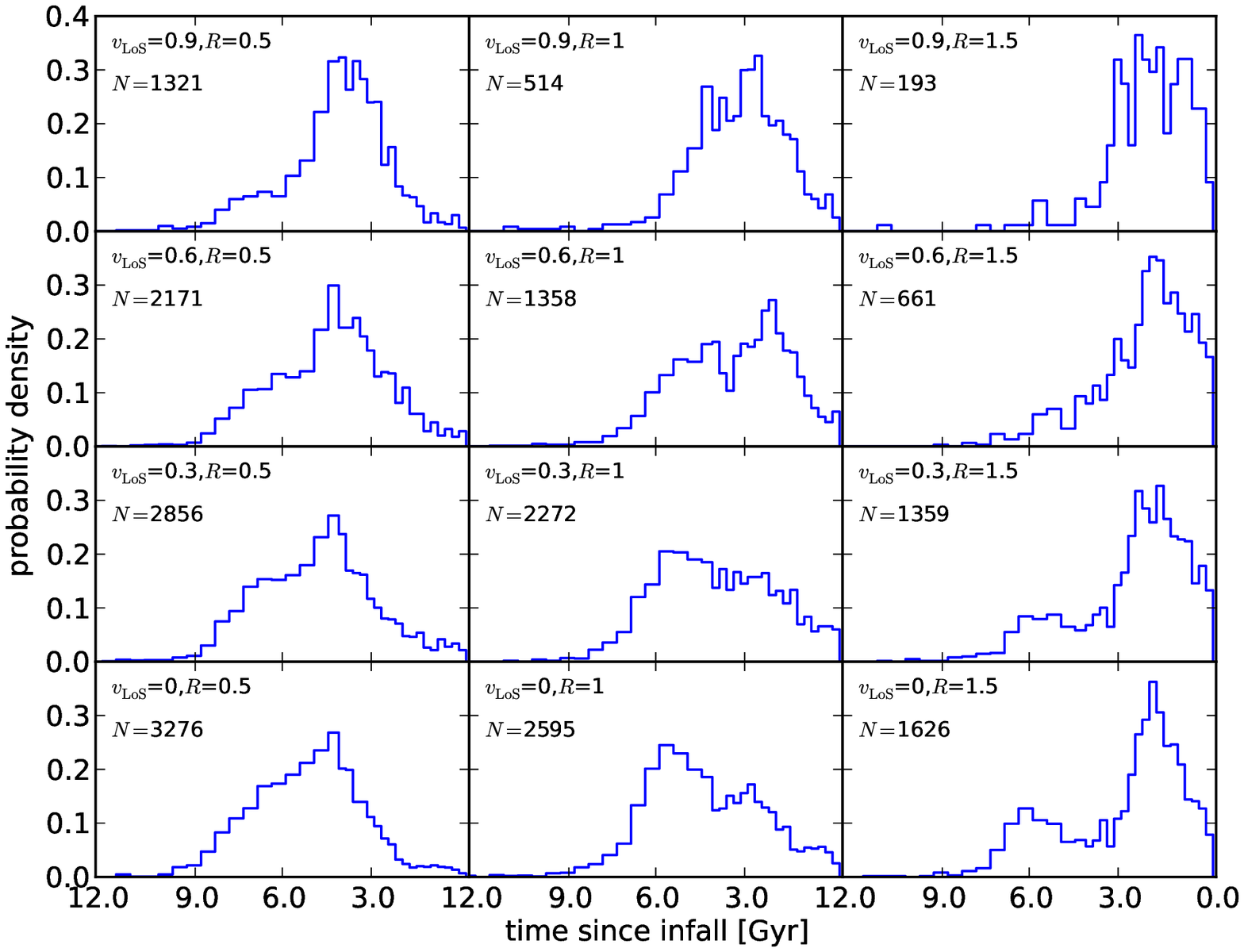}
\caption{Probability density functions (PDFs) of infall times for a selection of points $(R/r_{\textrm{vir}},V/\sigma)$ in projected phase space. Each point is sampled in a region out to a distance about 3 per cent of the full range of each coordinate from the labelled point. The number of satellites in each region is labelled $N$, giving a measure of the statistics of each PDF.\label{sample_pdfs}}
\end{figure*}

Since our ultimate goal is to model quenching, perhaps the most directly useful question we can ask of our dataset is, given a position in projected phase space, what is the distribution of possible infall times (or another parameter of interest), and what is the likelihood of each? This question is easily answered in a statistical manner by sampling all satellites in a small region of phase space and binning them by infall time. The result is shown in Fig.~\ref{sample_pdfs} for a selection of points in phase space, varying both the radial coordinate (left-right across the panels) and the velocity coordinate (up-down). We focus here on the trends with the velocity coordinate. The first trend we note is that at a given radius, satellites with higher $v_{\textrm{LoS}}$ typically have slightly more recent infall times than their low $v_{\textrm{LoS}}$ counterparts. This is because a satellite with high $v_{\textrm{LoS}}$ is more likely to have a high total speed and can penetrate deeper into the host potential in a given amount of time than a satellite with low speed.

In some cases, the velocity coordinate allows us to discriminate between different satellite populations. Consider the rightmost panels ($R=1.5$) of Fig.~\ref{sample_pdfs}. One peak in the distribution of infall times is clearly visible at all values of $v_{\textrm{LoS}}$, at $a \approx 0.85$; it is made up of infalling satellites. A second peak at $a \approx 0.6$ is made up of backsplash satellites, and is only present for low values of $v_{\textrm{LoS}}$. The backspash galaxies have lower kinetic energy relative to the host potential than recent infalls due to a combination of mass accretion by the host (increasing $\sigma$, therefore causing an apparent slowing of haloes) and dynamical friction. 

Given a sufficiently large sample of observed galaxies, the PDFs we have created will allow us to statistically assign an infall time to each one using all the available dynamical information and a meaningful uncertainty in our assignment. We have developed a code and a compact data table to this end which we are prepared to share on request.

\begin{figure}
\leavevmode \epsfxsize=\columnwidth \epsfbox{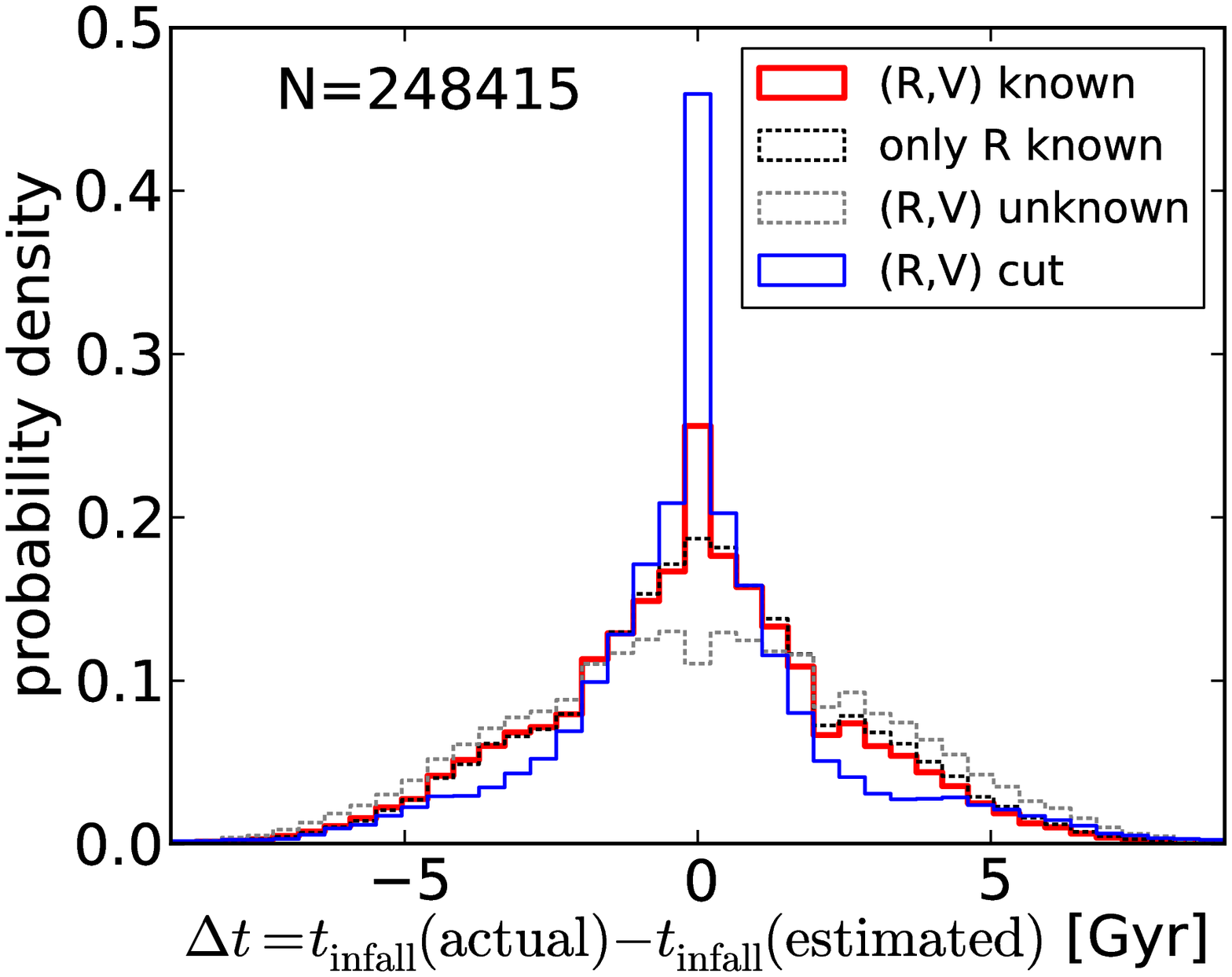}
\caption{The distribution of $\Delta t$, a measure of the rate of success of using our PDFs to estimate the infall time $t_{\textrm{infall}}$ of satellite haloes in our sample, is plotted with a thick red solid line. $\Delta t=0$ represents a successful estimation, while increasing $|\Delta t|$ represents increasing an increasing discrepancy with the correct value. A sharp peak at $\Delta t=0$ therefore represents a high rate of success. The black dotted line shows the distribution if we assume no knowledge of the LoS velocity of the satellites, while the gray dotted line shows the distribution if we assume no knowledge whatsoever of the coordinates of the satellites. Each additional coordinate used increases the reliability of the predictions. The thin solid red curve shows the distribution of $\Delta t$ for a subsample of satellites selected to study a particular population of satellites (see text for full details). This shows an example of restricting the region of phase space considered to increase the reliability in estimating $t_{\textrm{infall}}$. We quantify the degree of success in estimating $t_{\textrm{infall}}$ by the sharpness of the peak in the distribution of $\Delta t$, measured by its standard deviation. The standard deviations for the four curves are 2.58~Gyr, 2.64~Gyr, 3.10~Gyr, 2.48~Gyr (in the same order as labels in the legend). The number satellites used to test the PDFs is labelled $N$.\label{improvement}}
\end{figure}

To test the ability of our PDFs to correctly assign an infall time to a satellite halo, we use our PDFs to estimate the infall time of each satellite in our sample. This estimate is then compared to the actual infall time of the satellite, which we know from tracking its orbit. The distribution of $\Delta t = t_{\textrm{infall}}(\textrm{actual})-t_{\textrm{infall}}(\textrm{estimated})$ is plotted in Fig.~\ref{improvement}. A $\Delta t$ of 0 represents a correct guess, so a stronger peak about $\Delta t=0$ represents a higher rate of success. As a quantitative measure of the strength of the peak, we calculate the standard deviation of the distribution. By using our PDFs based on knowledge of the $(R,V)$ coordinates of each satellite, we assign the infall time correctly to within $\pm 2.58$~Gyr in 68 per cent of cases. For comparison, we also plot the distribution of $\Delta t$ obtained if only the position coordinate $R$ of the haloes is known, accurate to within $\pm 2.64$~Gyr in 68 per cent of cases, and if none of the coordinates are known; in this case, we simply draw values from the distribution of infall times of our entire sample at random, and find we are accurate within $\pm 3.10$~Gyr in 68 per cent of cases. Finally, we plot a curve showing the distribution of $\Delta t$ for a particular subsample of satellites chosen by eye which have $\frac{V}{\sigma}>-\frac{4}{3}\frac{R}{r_{\textrm{vir}}}+2$. This subsample is designed to retain primarily infalling satellites (see Fig. \ref{infall_map_proj_binned}), and in this case the assignment of infall times is even more reliable. This demonstrates that if a particular population of satellites is of interest, it is often possible to choose a region of phase space that maximises the likelihood of correctly assigning $t_{\textrm{infall}}$. For this example, we correctly assign $t_{\textrm{infall}}$ within $\pm 2.48$~Gyr in 68 per cent of cases. In practice, this makes it possible to increase the purity of an observational sample of galaxies at the expense of completeness of the sample by limiting the region of phase space from which the sample is drawn.

\begin{figure}
\leavevmode \epsfxsize=\columnwidth \epsfbox{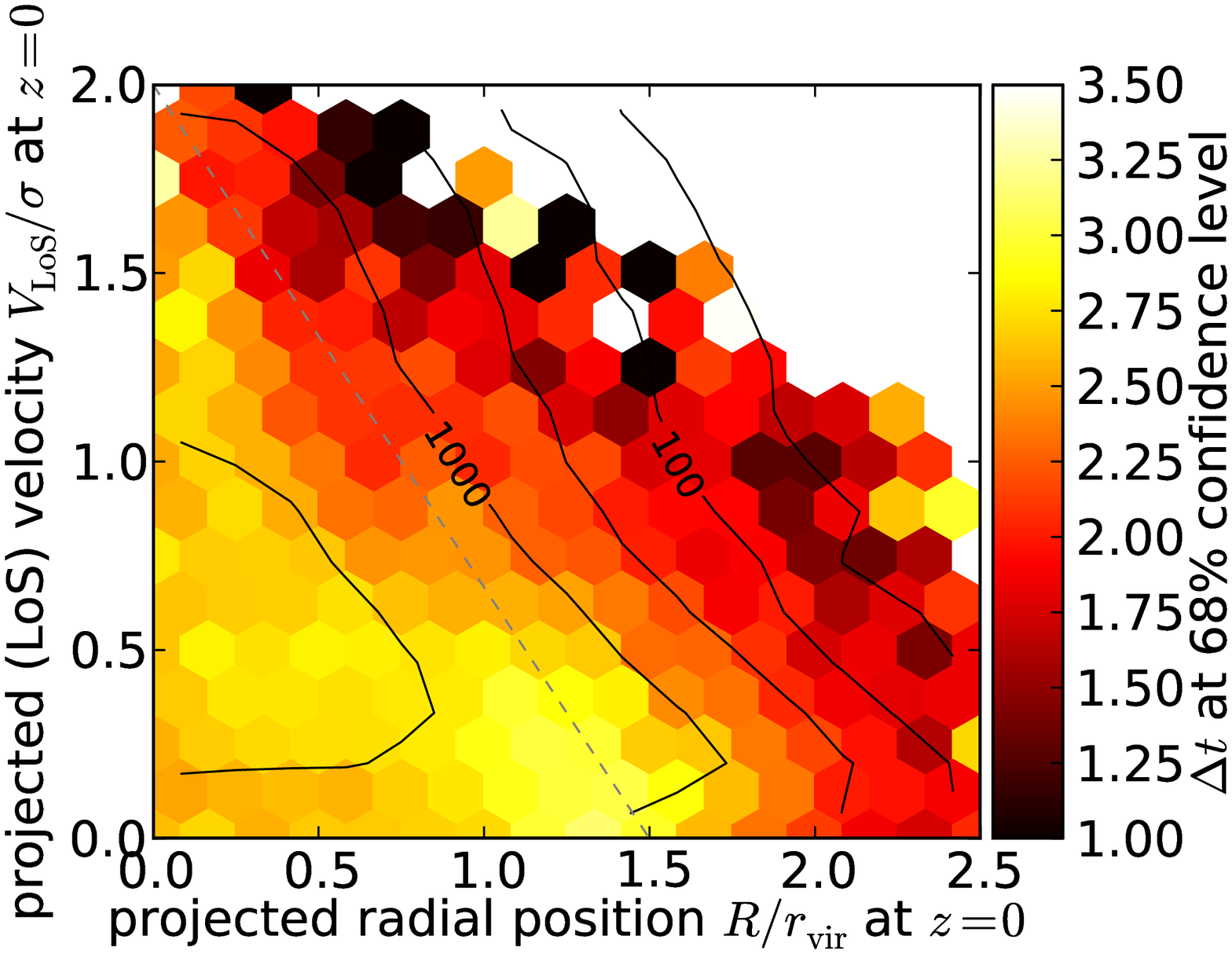}
\caption{The distribution of $\Delta t$ at the 68 per cent confidence level in the $(R,V)$ plane; our PDFs will be most accurate when applied to satellites that occupy the reddest (darkest) regions of the space. Use of the PDFs also requires a reasonable sample size in the region of interest; contours (0.5 dex intervals) illustrate the number of satellites per cell. We find that the PDFs become too sparse for reliable use in cells containing less than a few hundred satellites. The dashed gray line marks the same region as in Fig.~\ref{infall_map_proj_binned}\label{confidence_distribution}}
\end{figure}

Fig.~\ref{confidence_distribution} shows the accuracy of the assignment of $t_{\textrm{infall}}$ at the 68 per cent confidence level for bins in $(R,V)$ space. The PDFs give the most reliable results in those regions with both good accuracy and a large population of satellites. In the outskirts of clusters, an accuracy $\Delta t$ of about 3 Gyr is sufficient to robustly separate infalling and backsplash satellites.

We conclude that using all observable coordinates of a satellite $(R,V)$ offers an improvement in assigning $t_{\textrm{infall}}$ correctly over using only the position coordinate $R$ comparable to, but somewhat less than, the improvement seen when using the position coordinate $R$ rather than no knowledge of satellite coordinates at all. Careful selection of the region of phase space to be sampled can, in many cases, further increase the reliability in estimating $t_{\textrm{infall}}$.

\subsection{Mass trends}\label{subsec-mass}

While haloes of different masses are expected to be self-similar in most ways, we still find some trends with both host halo mass and satellite halo mass. We study these trends by separating the haloes of interest -- either hosts or satellites -- crudely into a `high mass' and a `low mass' bin, then compare the distribution of satellites in phase space and infall time for the two bins.

\begin{figure*}
\leavevmode \epsfxsize=\columnwidth \epsfbox{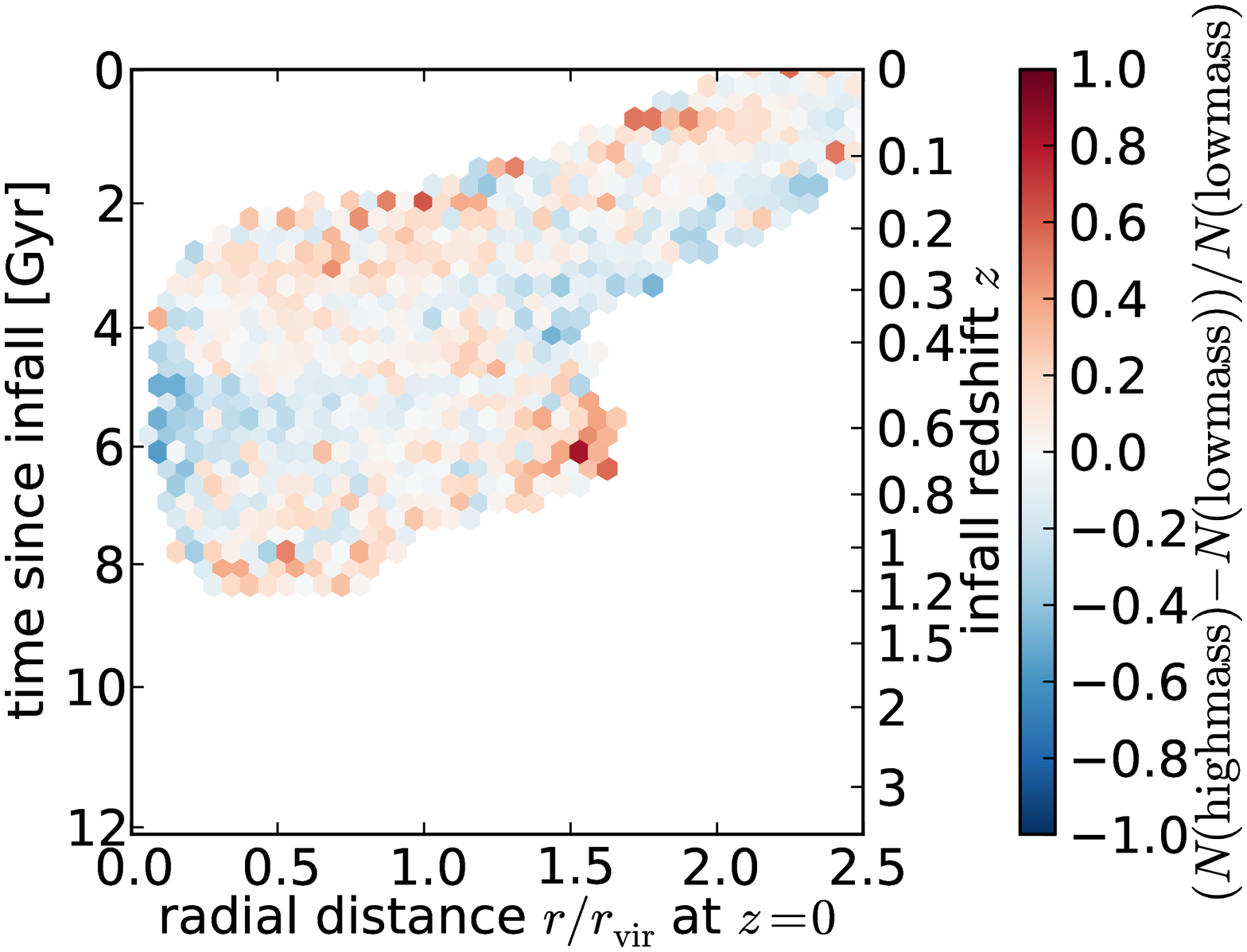}
\hfill
\leavevmode \epsfxsize=\columnwidth \epsfbox{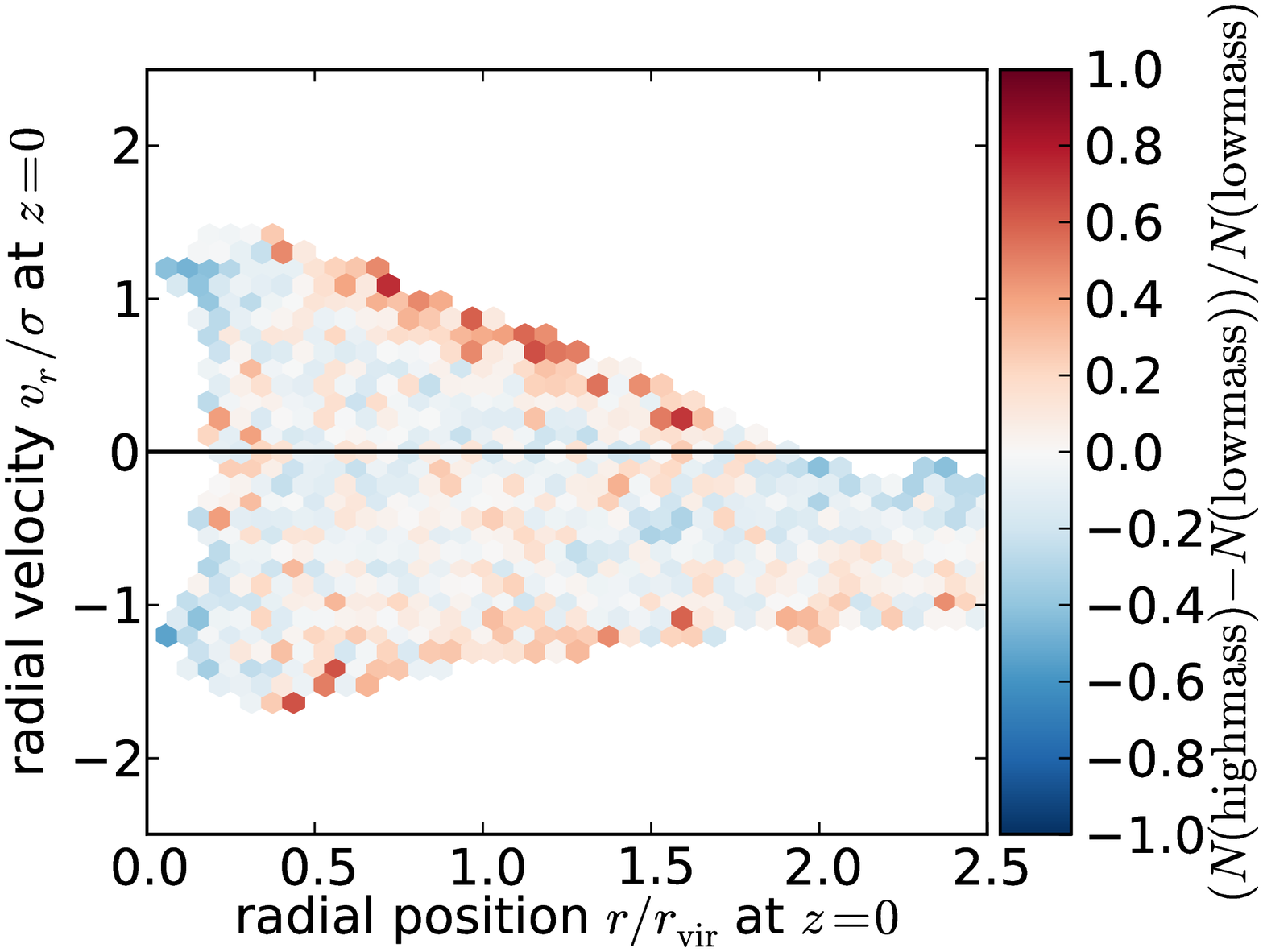}
\caption{Relative abundance of satellites in the space of infall times vs. radial positions (left panel) and phase space (right panel), comparing satellites around low mass ($10^{14}-10^{14.5}$~\Msun) hosts and high mass ($10^{14.5}-10^{15}$~\Msun) hosts. The two populations are normalized, subtracted in each bin, then compared to the fiducial abundance of satellites in that bin (arbitrarily chosen to be the population with low mass hosts). Only bins containing at least 100 satellites are shown.\label{host_infall}}
\end{figure*}

Fig.~\ref{host_infall} shows a comparison of satellites around high- and low-mass hosts in infall time vs. radial position space (left panel), and in phase space (right panel). Low mass hosts are defined to have $10^{14}$~\Msun~$< M < 10^{14.5}$~\Msun while high mass hosts are in the range $10^{14.5}$~\Msun~$< M < 10^{15}$~\Msun. The population of satellites around each type of host is normalized, then the two are subtracted in each infall time - radial distance bin and compared to the fiducial abundance of satellites in that bin. Cells with a reddish colouration are preferentially occupied by satellites around high mass hosts, while cells with a bluish colouration are preferentially occupied by satellites around low mass hosts.

In the infalling population of satellites ($t<4$~Gyr in Fig.~\ref{host_infall}, left panel) the satellites of high mass hosts appear to fall into their host somewhat faster than the satellites of low mass hosts; at a given time since infall, a satellite of a low mass host has a larger typical radial distance than a satellite of a low mass host. We propose two possible explanations for this effect. First, satellites of high mass hosts tend to have more radial orbits than satellites of low mass hosts \citep{2011MNRAS.412...49W}, causing them to move radially inward more rapidly on average. Alternately, because the hosts are continually accreting mass, their virial radius gradually increases. Hosts with higher masses have higher present-day accretion rates; this is visible in Fig.~\ref{host_infall} (right panel) where the region of phase space occupied by infalling satellites shows an excess of satellites around higher mass haloes \citep[see also][fig. 3 therein especially]{2002ApJ...568...52W}. Their satellites therefore appear to move inward slightly faster than those of low mass hosts in this coordinate system. These two explanations could be considered an argument for a different normalization of the radial coordinate, however other choices (e.g.\ the virial radius of the host at the infall time of each satellite) still introduce similar effects, and the present virial radius of the host has the advantage of being more related to observable quantities. A more practical approach is to compute PDFs for narrow host mass bins ($\sim 0.5$ dex) and use whichever is applicable to an observed system of interest; our sample contains enough satellites to make this feasible.

\begin{figure}
\leavevmode \epsfxsize=\columnwidth \epsfbox{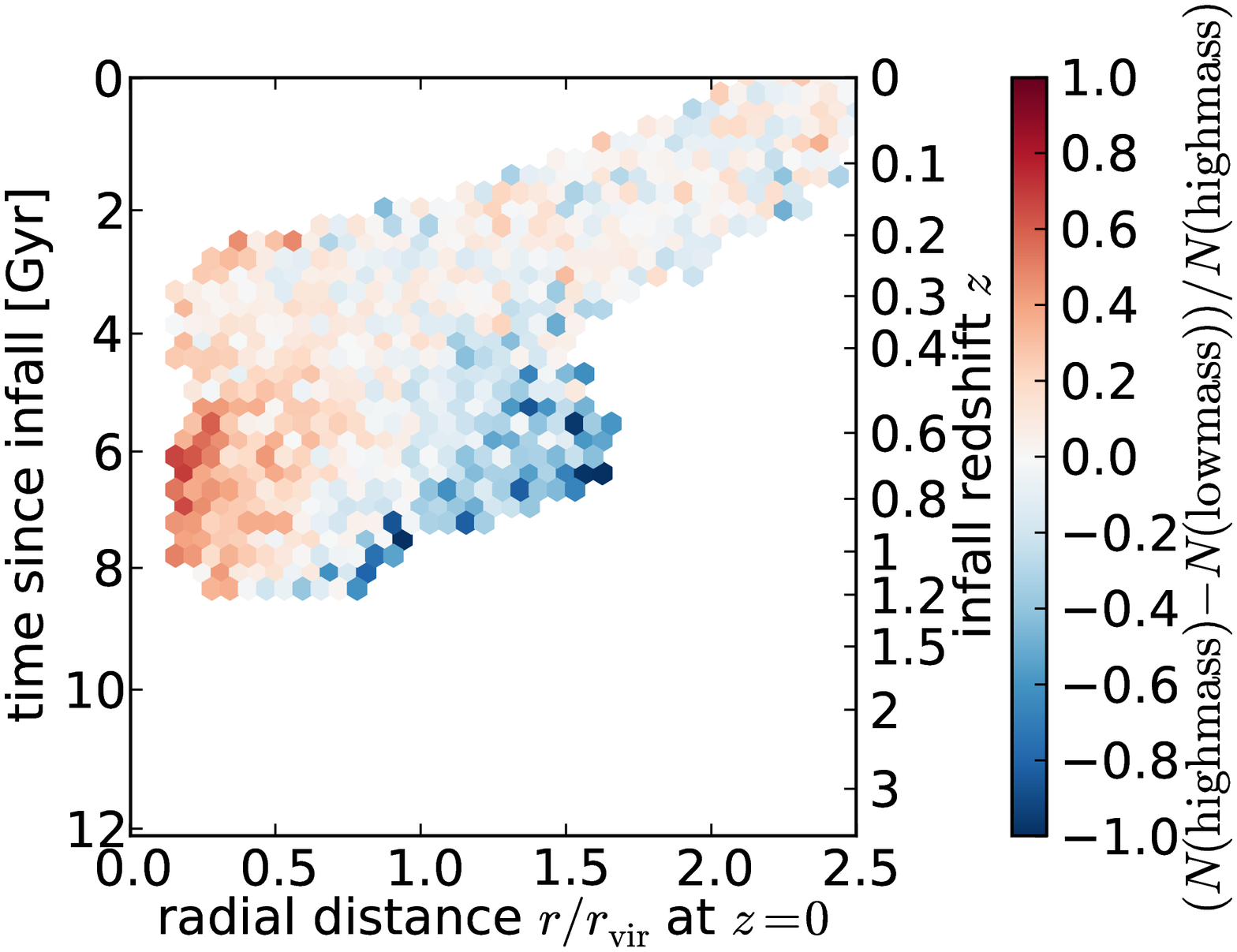}
\caption{Relative abundance of satellites in the space of infall times vs. radial positions comparing low mass ($10^{11.9}-10^{12.2}$~\Msun) and high mass ($10^{12.2}-10^{12.8}h^{-1}$~\Msun) satellites. The two populations are normalized, subtracted in each bin, then compared to the fiducial abundance of satellites in that bin (arbitrarily chosen to be the high mass population). Only bins containing at least 100 satellites are shown. High mass satellites are slowed more than low mass satellites by dynamical friction and therefore have lower typical backsplash distances than their low mass counterparts.\label{sat_infall}}
\end{figure}

We also consider trends with satellite mass. Because satellite mass typically decreases with increased time spent in a cluster environment, we use the satellite mass at time of infall (first crossing of $2.5$~$r_{\textrm{vir}}$) as a characteristic mass for this analysis. Fig.~\ref{sat_infall} shows that lower mass satellites experience backsplashes to larger radii than high mass satellites. This is because the slowing due to dynamical friction is proportional to the mass of the satellite, so that higher mass satellites lose a larger fraction of their kinetic energy during a passage through a cluster. To account for this trend with satellite mass, separate PDFs can be produced for relatively narrow bins in satellite mass.

Care must be taken when using our method to interpret the distribution of galaxies in a cluster. Baryons (especially the stellar component) are typically more tightly bound than their associated dark matter halo, and should therefore outlive the halo in the cluster environment. This restricts the satellites, as traced by galaxies, that can be studied using our method to those which we expect to have a surviving DM halo. With the MDR1 dataset, this corresponds to galaxies with associated DM halo masses of $\sim 10^{11.9}$~\Msun or greater. Smaller satellites may still be studied by using higher resolution simulations (e.g.\ the Bolshoi simulation of \citet{2011ApJ...740..102K}, which has identical parameters to MDR1 but with higher resolution and a smaller box size). Low mass satellites are much more abundant, so the smaller volume of a higher resolution simulation should not impact our ability to obtain a large sample, except perhaps for low mass satellites around very large hosts (of which there may only be a few in a simulation such as Bolshoi). 

\subsection{Resolution Effects}\label{subsec-resolution}

As shown in section \ref{sec-method}, some halos which experience a close approach to the centre of a larger halo experience artificial disruption. We estimate that any given bin in our PDFs is less than 20\% incomplete, and that our sample of satellites as a whole is less than 4\% incomplete down to $10^{11.9}$~\Msun. The effect on the predictive power of our method was investigated by generating an orbit catalogue with extra orbits added to compensate for artificially disrupted halos. The effect of artificial disruption on the distributions of $\Delta t$ (such as those shown in Fig. \ref{improvement}) is very small, causing changes on the order of a few tenths of a percent.

Our initial interest for applications of our method is to large satellites (halo masses $\gtrsim 10^{12.5}$~\Msun) around hosts with halo masses $\gtrsim 10^{14}$~\Msun, which motivated our choice of the MDR1 simulation as a starting point. Applying our method to smaller satellites of smaller hosts is simply a matter of choosing an appropriate simulation; our method is easily applied to any simulation that can be processed by the \textsc{rockstar} code. The Bolshoi simulation is of particular interest since it is identical to MDR1 in all respects save resolution and could therefore be used in conjunction with MDR1 to extend the range of our method to lower mass ranges while maintaining good statistics at the high mass end.

\section{Conclusions}\label{sec-conclusions}

The phase space distribution of infalling, backsplash and virialized satellite halos are different but not everywhere distinct. The LOS velocity distributions we recover are in agreement with the results of \citet{2005MNRAS.356.1327G}. Like them, we find that different populations of satellite halos are better separated in phase space than in the radial coordinate alone, but that there is no immediately obvious cut that we can impose on the projected phase space coordinates of a satellite galaxy to separate different populations. \citet{2011MNRAS.416.2882M} identified some regions of projected phase space where parts of the infalling or backsplash populations could be picked out with little contamination by other populations. Building on this idea, we have examined the entire projected phase space and determined the confidence with which the time since infall can be assigned to a satellite occupying that region. This lets us easily identify regions where the time since infall can be determined accurately enough to reliably separate satellite populations. Such regions that we identify are a superset of those identified by \citet{2011MNRAS.416.2882M}. This allows an increase of the area of projected phase space and therefore the number of satellite galaxies available for use in studying star formation and other effects that may depend on the orbital history of a satellite.

We have developed a tool that estimates the infall time and a confidence in the estimate for a satellite halo given its projected phase space coordinates. The possible infall times given a pair of projected phase space coordinates are weighted according to the frequency of their occurence in simulation, then an infall time is randomly selected from this weighted distribution. We predict that, when applied to a large sample of observed galaxies, this method will allow correlations between infall times and satellite star formation histories to be studied. Our method is easily adapted to other similar parameters (closest approach to host centre, time since first pericentre, etc.) that will be considered in our forthcoming study of SF quenching in cluster environments, and to `preprocessing' scenarios -- accretion on to a group sized halo before accretion on to a cluster sized one. It would be straightforward to adapt our method to systems at higher redshift where we think the different satellite populations may be better separated than at low redshift. Our framework would also be very well suited to studying dark matter stripping of satellites.

\section*{Acknowledgements}\label{sec-awknowledgements}

We wish to thank the authors of the Bolshoi and MDR1 simulations for making their simulation outputs publicly available.

PSB was supported by a Hubble Theory grant; program number HST-AR-12159.01-A was provided by NASA through a grant from the Space Telescope Science Institute, which is operated by the Association of Universities for Research in Astronomy, Incorporated, under NASA contract NAS5-26555.

MJH acknowledges support from an NSERC Discovery grant.

\bibliography{reflist}

\label{lastpage}

\end{document}